\input{aipcheck}

\newcommand{\HESS}{H.E.S.S.}

\newcommand{\naive}{na\"{\i}ve}

\documentclass[
    ,final            
  ]
  {aipproc}

\layoutstyle{8x11double}

\begin{document}

\title{Future science issues for Galactic \\ very-high-energy gamma-ray astronomy}

\classification{98.38.-j,98.20.-d, 97.60.Jd, 97.60.Lf}
\keywords      {gamma-rays: observations, gamma-rays: theory}

\author{Diego F. Torres}{
  address={  ICREA \& Institut de Ciencies de l'Espai (IEEC-CSIC),
  Campus UAB, Fac. de Ciencies, Torre C5, parell, 2a planta, 08193
  Barcelona,  Spain.  }
}

\begin{abstract}
This work intends to provide a brief summary of some of the Galactic science issues for the next generation of very high energy (VHE) instruments. The latter is here generically understood, as an instrument or set of instruments providing about one order of magnitude more sensitivity at its central energy
(at about 1 TeV), but extending the observational window to have a real broadband capability (from a few tens of GeV up to tens of TeV); exceeding at low energies the current VHE threshold for observations set by MAGIC as well as the few-tens-of-GeV sensitivity set by Fermi. 
Science topics regarding populations of emitters, pulsars and their nebula, binaries, supernova remnants, stars, and their associations, are discussed. 
\end{abstract}

\maketitle

\section{Introduction}

This work intends to provide a brief summary of some of the most important Galactic science issues that will focus our attention provided a new ground-based array, more sensitive, with lower energy threshold, and enhanced angular resolution as compared with the currently operating instruments, is available. One such forthcoming instrument may be the {\it Cerenkov Telescope Array}, or in short CTA, for which many European institutions have already joined efforts to work in a Design Study. Other papers in this volume (e.g., by J. Hinton, or M. Martinez) will give account of the current status of different aspects of this design.  
This paper is then not about on our ideas for CTA, nor it is a review of TeV astronomy. Because of that we advice the reader not to expect 
a thorough coverage of all current TeV results in the topics mentioned below. Rather, we will mention some previous results as a path to present ideas 
that could be useful in defining {\it some} of these future Galactic science issues; physics we know is there for us to test, but that still escape the reach of current instrumentation.

\section{GeV to TeV surveys: distinguishing populations}

One of the first issues for Galactic science with the new generation of instruments will probably be to achieve
a distinction, at a level of population studies, of the classes of emitters in the different energy domains.
22 VHE gamma-ray sources have
been reported in the 2004 and 2005 \HESS\ survey of the inner Galaxy (GPS). The third EGRET 
catalogue (Hartman et al. 1999, soon to be superseded by the advent of the new Fermi catalog!)
still
represents the companion to these VHE sources
 above an energy threshold of 100 MeV (with peak sensitivity
between 150 and 400 MeV). It lists 271 sources, 17 of which are located within the
\HESS\ GPS region. 
Comparing the
instrumental parameters of VHE instruments and EGRET there is a clear
mismatch both in angular resolution and in sensitivity: In a $\sim 5$ hour observation (as a
typical value in the GPS region) \HESS\ is a factor of $\sim 50-80$
more sensitive (in terms of energy flux $E^2 dN/dE$) than EGRET above
1~GeV in the Galactic Plane for the exposure accumulated in the third EGRET catalogue. Assuming a
similar energy flux output in the two different bands this mismatch
implies at first sight that \HESS\ sources are not likely to be
visible in the EGRET data set. Conversely (again under the assumption
of equal energy flux output), VHE gamma-ray instruments should be
able to detect the majority of the EGRET sources, as has been
suggested in the past. 

\begin{figure}
    \includegraphics[height=.2\textheight]{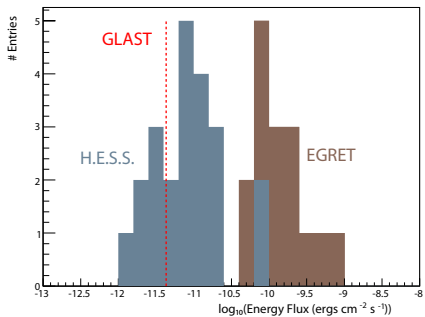} 
\caption{Distribution of integrated energy flux $\nu F \nu$ for
  sources in the Inner Galaxy. For EGRET, the energy
  flux between 1~GeV and 10~GeV, for the \HESS\ sources, the energy
  flux between 1~TeV and 10~TeV, are shown. Also shown is the
  sensitivity prediction for the GLAST-LAT for a typical location in
  the Inner Galaxy (l=10, b=0). From Funk et al.  (2008).}\label{fig::EnergyFluxDistribution}
\end{figure}

Figure~\ref{fig::EnergyFluxDistribution}
compares the energy fluxes $\nu F \nu$ for EGRET sources and \HESS\
sources in the inner Galaxy. Clearly, the EGRET sources do not reach
down as low in energy flux as the \HESS\ sources, a picture that should
change with Fermi (referred to as GLAST in the Figure). 
By comparing the two surveys and analyzing the matching of the spectral properties in both energy regimes, it was found that (Funk et al. 2008)
\begin{itemize}
\item There are rather few spatially coincident GeV-TeV sources for
  the considered Galactic region, and they could occur by
  chance, the chance probability of detecting two coincident sources
  within the \HESS\ GPS region is $\sim 40$\%, thus no strong hint for
  a common GeV/TeV source population is detected.
  \item Spectral compatibility (based on a power-law extrapolation) seems
  present for most of the positionally coincident sources, but again,
  this is expected to occur by chance

\item  \HESS\ limits at the position of the EGRET sources are
  constraining for a power-law extrapolation from the GeV to the TeV
  range for several of the EGRET sources, strongly suggesting cutoffs in
  the energy spectra of these EGRET sources in the unexplored region
  below 100~GeV. Power-law extrapolation of EGRET spectra seem to be
  ruled out for most of the EGRET sources. On the contrary,
EGRET limits at the position of the \HESS\ sources are
  not constraining for a power-law extrapolation from the TeV to the
  GeV range. This picture may change for Fermi, which upper limits, if so are imposed, will be
  severely restricting the SEDs.
      
\end{itemize}

But in addition of these facts, even from theory along the expectation of equal
energy flux in the GeV and TeV band (leading to the same populations in both bands) is \naive\ and can easily be wrong: 
EGRET sources may
not emit comparable energy fluxes in the VHE $\gamma$-ray band but
exhibit cut-offs or breaks in the energy band between
EGRET and \HESS. There are well motivated
  physical reasons why the population of GeV and of TeV sources might
  be distinct: acceleration limits, particle transport affecting the primary spectrum of cosmic rays, convection, particle energy losses producing steepening, particle injection limited in time, absorption by pair production, redistribution of photon energies by cascading process.
Additionally, TeV instruments are typically only sensitive to
emission on scales smaller than $\sim 1^{\circ}$. If any of the EGRET
sources are extended beyond $1^{\circ}$ without significant
sub-structure on smaller scales (not precluded given the poor
angular resolution of EGRET), current Imaging Cherenkov instruments may not be
able to detect them since these sources would completely fill the
field of view (FoV) and be removed by typical background subtraction
methods. 
A broadband instrument covering from tens of GeV to TeV with above Fermi sensitivity
in the corresponding band would allow to distinguish the different populations present at the high energy sky, separating the technical from the physical reasons by which these may diverge.

\section{Pulsars}

Pulsar science cases for VHE observatories have been recently rekindled by the MAGIC detection of pulsed emission above 25 GeV coming from Crab, revealing a relatively high energy cut-off in the phase-averaged spectrum (Aliu et al. 2008, see the specific article presented in this volume with details of  this measurement). This high energy cutoff probably indicates that emission happens far out in the magnetosphere where the influence of the magnetic field, producing absorption of the form $\gamma + B \rightarrow e^\pm$, is not enough to lead to a super-exponential cutoff of the high energy emission.

Certainly, one of the fundamental unanswered questions about pulsars concerns the mechanism and location by which GeV (and TeV) emission is radiated. Future sensitive arrays with low energy threshold could test this question with a large sample since 
essentially,  pulsed detection is limited by photon statistics. Fermi, about 100 times more sensitive than EGRET above 10 GeV, would allow  detections more than 25 times fainter, or more  than 5 times more distant (reaching to the Galactic center). 
However, whereas Fermi sensitivity worsens with energy in the range $E>10$ GeV, any sensitivity  of future arrays will significantly improve. To observationally study the pulsar cutoffs with a large sample, and to make a significant improvement in population studies (e.g., what
happens as the observed gamma-ray pulsar luminosity --seemingly proportional to $\dot E^{0.65}$-- approaches the total available spin-down luminosity $\dot E$? how much does the assumption of a 1 sr beaming solid angle distort this correlation? how does this correlation look like above 10 GeV?) is probably a mission for future ground-based arrays only. 

Especially at high energies, many Fermi  sources will be unidentified and future arrays could search, blindly, for the appearance of pulsar periods.  
We recall here that for pulsars, the phase changes as  $\phi=2\pi (t \nu + t^2 \dot \nu)$,  with  $t$ being the observation time and $ \nu=1/P$, with $P$ the period and $\dot P$ the period derivative. The variation of phase due to change in the period can be neglected if its phase contribution is at most, less than half cycle, i.e., $ t^2 \dot \nu < 1/2 $.   Thus, the duration of the time for the search in absence of a priori information about $P$ and $\dot P$ should satisfy $T_{\rm obs} < (1/ 2 \dot \nu)^{0.5}$. During this time, enough photons should be collected to determine $P$. To give an example, the useful blind observation time for Crab is 10 hours. Thus, Fermi can only deal with the brightest pulsars without a-priori determined timing solutions. 
Instead, Fermi high energy sources should all be bright detections for future VHE arrays. Will this provide a way of measuring 
the fraction of radio quiet gamma-ray pulsares directly? 

Future ground-base arrays can help complete the pulsar panorama by studying extreme cases: pulsars with very high and very low magnetic fields, i.e., millisecond pulsars and magnetars. In polar cap models, the high energy cutoff of millisecond pulsars is a decreasing function of the magnetic field, so that pulsars with $10^8$ to $10^9$ G could present cutoffs around or above 100 GeV, well in the range of future arrays, and their observations could constitute a direct test of the polar cap theory. On the other hand, the high energy emission from magnetars is also probably beyond the Fermi capabilities, as exemplified directly in Fig. 8 of Zhang \& Cheng (2002, also 2001). Soft gamma-ray repeaters and anomalous X-ray pulsars are magnetars, presenting periods $P = 6 - 12$ s, being persistent X-ray sources (much beyond spin down) with luminosities of the order $10^{34}$ -- $10^{36}$ erg s$^{-1}$, and dipole fields $10^{14}$ -- $10^{15}$ G in neutron stars. AXPs would be the first class of astrophysical objects with emission driven by magnetic field decay (Thomson \& Duncan 
1996). For magnetars, we do not expect polar cap emission: due to the large magnetic field all high energy photons would be absorbed, i.e., in these cases, future arrays would be testing outer gaps models. 

Finally, what about GeV-TeV emission related with short-timescale pulsar phenomena?
Buried beyond the current reach of working instruments, future arrays could explore any possible high-energy phenomenology related with timing noise, in which the pulse phase and/or frequency of radio pulses drift stochastically, or with another type of irregularity in the radio called glitches, a sudden increase in the pulse frequency produced by apparent  changes in the momentum of inertia of neutron stars.

\section{SNRs and cosmic rays}

\subsection{Brief general considerations on current SNR observations: what' next?}

Measurements from current IACTs has shown that a few SNRs  are particles accelerators up to energies above a few hundreds of TeV. We will exemplify with one case: for 
RX J1713.7-3946  (Aharonian et al. 2006a), photons  up to energies $ E>40$ TeV have been detected. This
provides proof of particle acceleration in the shell of RX J1713.7-3946 beyond 10$^{14}$ eV, up to energies close to the knee in the cosmic ray spectrum. Indeed, if
VHE gamma rays are produced via pion decay following inelastic proton-proton interactions, gamma-ray energies of $\sim$30 TeV 
imply that primary protons are accelerated to 
$\sim$ 200 TeV in the shell of RX J1713.7-3946. On the other hand, if 
the gamma rays are due to Inverse Compton scattering of VHE 
electrons, accelerated in the shell, off Cosmic Microwave Background photons (neglecting the presumably small contributions from starlight and infrared photons), the electron energies at the current epoch can be estimated in the Thompson regime as 
$\sim$ 110 TeV. Even considering Klein-Nishina effects $E_e \sim $ 100 TeV would be a realistic 
estimate. But, is a final proof for the gamma-ray origin (mostly leptonic or hadronic) in this and other sources only to come with future sensitivity, broadband coverage arrays?
 It must be recalled that Fermi, which
would 
need about 5 years of integration (with optimistic treatment of the diffuse GeV background) for a sensible distinction between the simplest hadronic and leptonic models above 1 GeV  for RX J1713.7-3946 (which is in addition particularly difficult for Fermi since there is an EGRET source 5 times more luminous nearby, Funk et al. 2008), would not have angular resolution or sensitivity enough to do this for most other (all others?) SNRs, and it will certainly be very difficult for it to determine space dependent changes in the spectrum of the emitted gamma-rays in the way it has been already determined by current instruments.

\subsection{Observing non-linearity in shocks}

Fermi 1st order acceleration is probably 
not a good enough description of particle behavior in SNR shocks.
In Fermi theory, typical power law spectra with 
slope $\sim 2-2.4$ are generated as a result of the acceleration process. 
But even when a small fraction of particles are injected in the 
acceleration box, they can take away an appreciable 
fraction of the available energy, making the dynamical reaction 
of the accelerated particles non-negligible (for a review see Malkov \& 
Drury 2001). If so, the theory of diffusive particle acceleration at shocks
becomes non-linear, predicting a high acceleration efficiency, a magnetic
field amplification, and particle spectra that is no longer described simply by power-laws.
Future broadband observations of SNRs could find these effects.

\subsection{Diffusion}

Perhaps one of the secure science cases for cosmic-ray physics with future arrays involve the study of diffusion. While travelling from the accelerator to the target the spectrum of cosmic rays is a strong function of time $ t$, distance to the source $R$, and the (energy-dependent)  diffusion  coefficient $D(E)$ (e.g., Aharonian \& Atoyan 2006). Depending on these parameters $t, R, D(E)$ one may expect any proton, and therefore gamma-ray spectrum: hard/soft/with and without TeV tail/with and without GeV counterpart, etc., and it gets even worse if the target is moving, like in the case of a stellar wind, where the process of convection can also play a role. Future instruments will open the parameter space in these three quantities, which is mostly unaccessible to current experiments. A determination, with high sensitivity, of spatially differentiated gamma-ray sources related to the same accelerator would lead to the experimental determination of the local diffusion coefficient $D(E)$ and/or of the local injection spectrum of cosmic rays (see, e.g., Gabici and Aharonian 2007).

\begin{figure}
\includegraphics[width=\columnwidth]{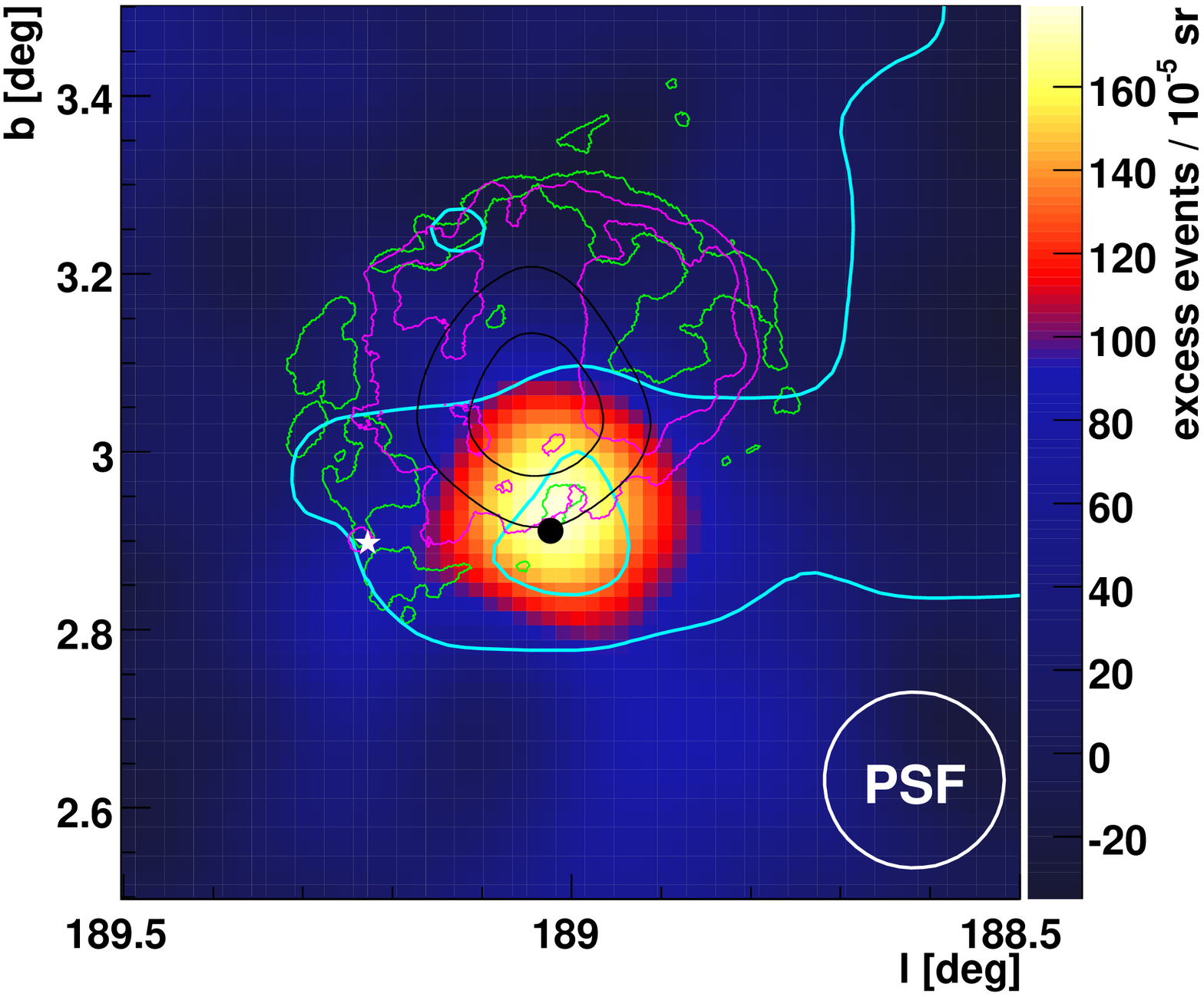}
\includegraphics[width=0.9\columnwidth, height=0.86\columnwidth]{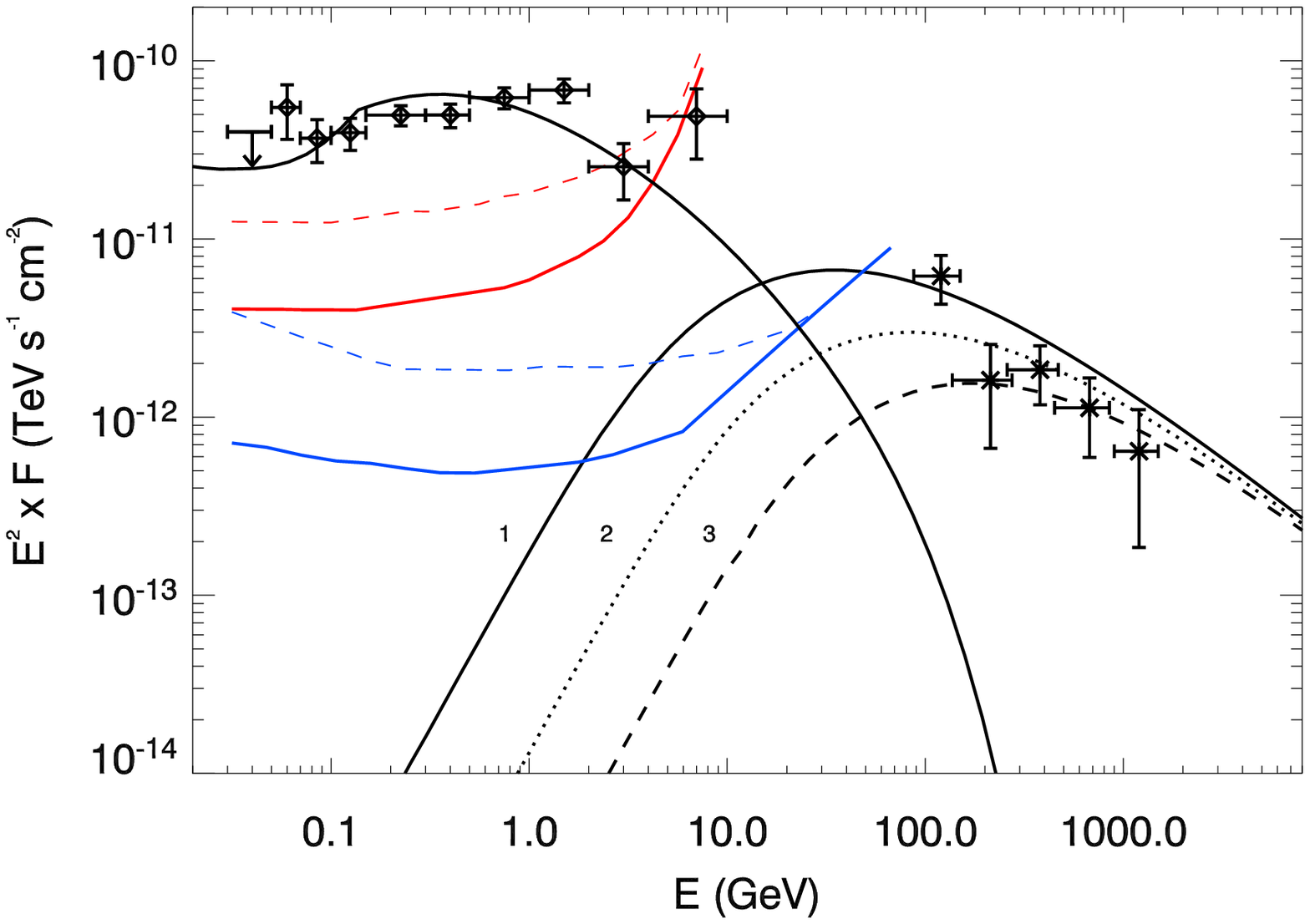}
\caption{Left: Sky map of $\gamma$-ray candidate events (background
subtracted) in the direction of MAGIC~J0616+225 for an energy threshold
of about 150~GeV in galactic coordinates. 
Overlayed are $^{12}$CO  emission contours
(cyan), contours of 20 cm VLA radio data
from  (green), X-ray contours from Rosat  (purple) and $\gamma$-ray contours from EGRET  (black).
The $^{12}$CO contours are at
7 and 14 K km/s, integrated from -20 to 20 km/s in velocity, the
range that best defines the molecular cloud associated with IC~443.
The contours of the radio emission are at
5 mJy/beam, chosen for best showing
both the SNR IC~443. 
The X-ray contours are at 700 and 1200 counts / $6 \cdot 10^{-7}$ sr. The EGRET contours represent a 68\% and 95\% statistical probability that a single source lies within the given contour.
The white star denotes the position of the pulsar CXOU J061705.3+222127. The black dot shows the position of the 1720 MHz OH maser . The white circle shows the MAGIC PSF of $\sigma = 0.1^{\circ}$. From Albert et al. (2007b), see also that paper for references to the multi-frequency data.
Right: MAGIC and EGRET measurement of the neighborhood of IC 443 (stars and squares, respectively) as compared with predictions of a model where cosmic-ray diffuse away from the 30 kyr SNR IC 443. 
At the MAGIC energy range, the curves show the predictions for a cloud of 8000 M$_\odot$ located at 20 (1), 25 (2), and 30 (3) pc, whereas they correspond to 
15 (1), 20 (2), 25 (3), and 30 (4) pc in the bottom panel. At the EGRET energy range, the curve shows the prediction for a few hundred M$_\odot$ located at 3--4 pc.
The EGRET sensitivity curve (in red) is shown for the whole lifetime of the mission 
for the Galactic anti-centre (solid), which received the largest exposure time 
and has a lower level of diffuse $\gamma$-ray emission, and 
for a typical position in the Inner Galaxy, 
dominated by diffuse $\gamma$-ray background. 
The Femi sensitivity curve (in blue) (taken 
from http://www-glast.slac.stanford.edu/software/IS/glast\_latperformance.html)
show the  1-year sky-survey sensitivity for the same position in the Inner Galaxy. From Torres et al. (2008).}
\label{IC443}
\end{figure}

Perhaps one of the best candidates for a diffusion of cosmic-rays scenario observed already with the current generation of instruments is IC443. Discovered by MAGIC (Albert et al. 2007b), the steep-spectrum VHE source is spatially dislocated from a nearby GeV detection.
The distribution of  molecular clouds and the multi-frequency phenomenology are consistent with  the interpretation of cosmic-ray interactions with a giant molecular cloud lying in front of the remnant, producing VHE gamma-rays but no counterpart at lower energies. At a morphological level, then, the lower the energy, the more coincident with the SNR the radiation should be detected. At a spectral level,  sufficient statistics should show that the lower the gamma-ray energy the harder the spectrum is (it is $\sim -2$ in the EGRET range). Both predictions should show in  future  observations, and with as a combination of GeV and TeV data. Figure \ref{IC443} provides a summary of this case. It is a clear tip of the iceberg for cases using future arrays, which broadband expected capability, and enhanced sensitivity will provide an arena for immediate testing of these ideas.

\begin{figure*}
\includegraphics[width=.71\columnwidth,trim=0 5 0 10]{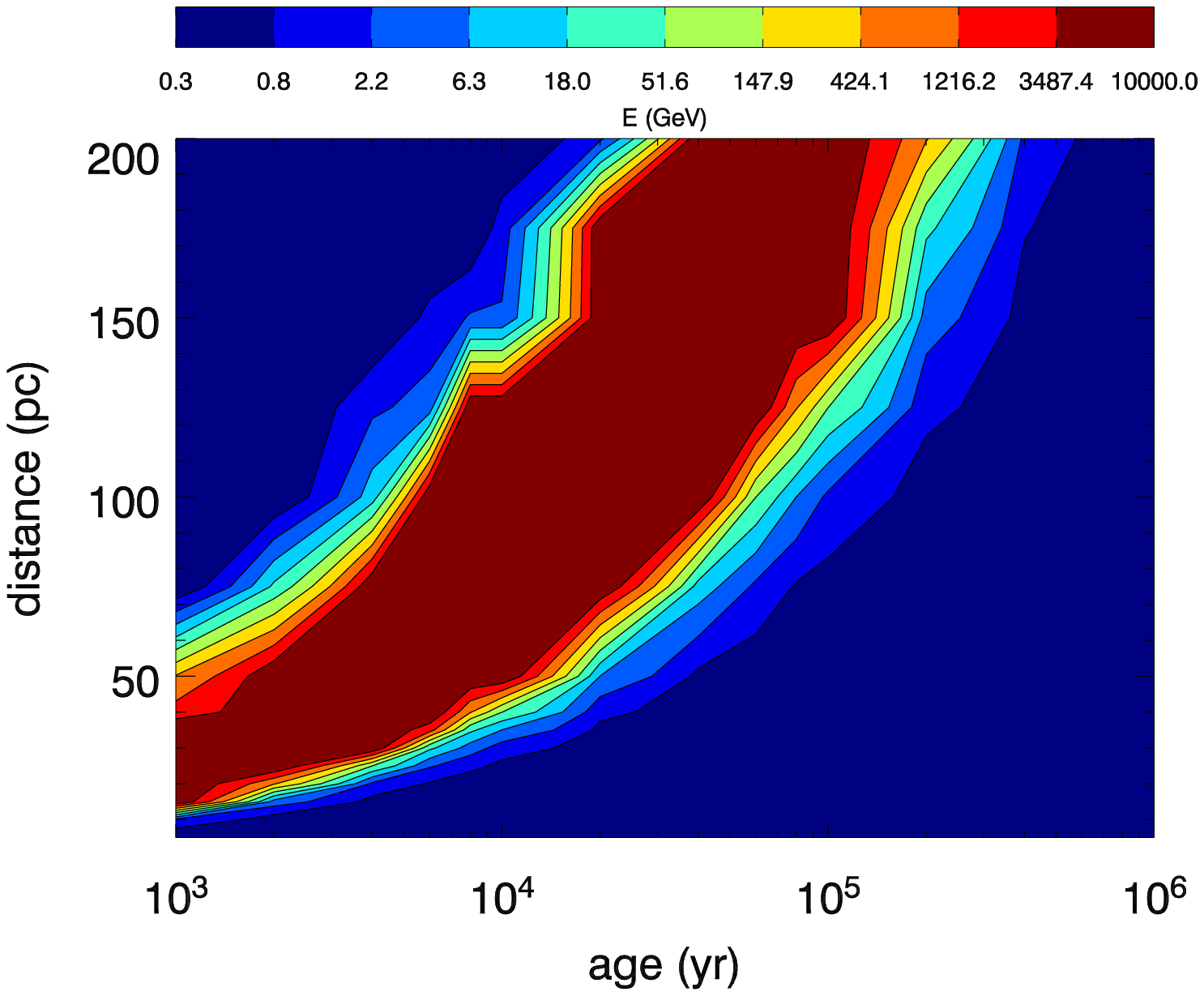}
\includegraphics[width=.71\columnwidth,trim=0 5 0 10]{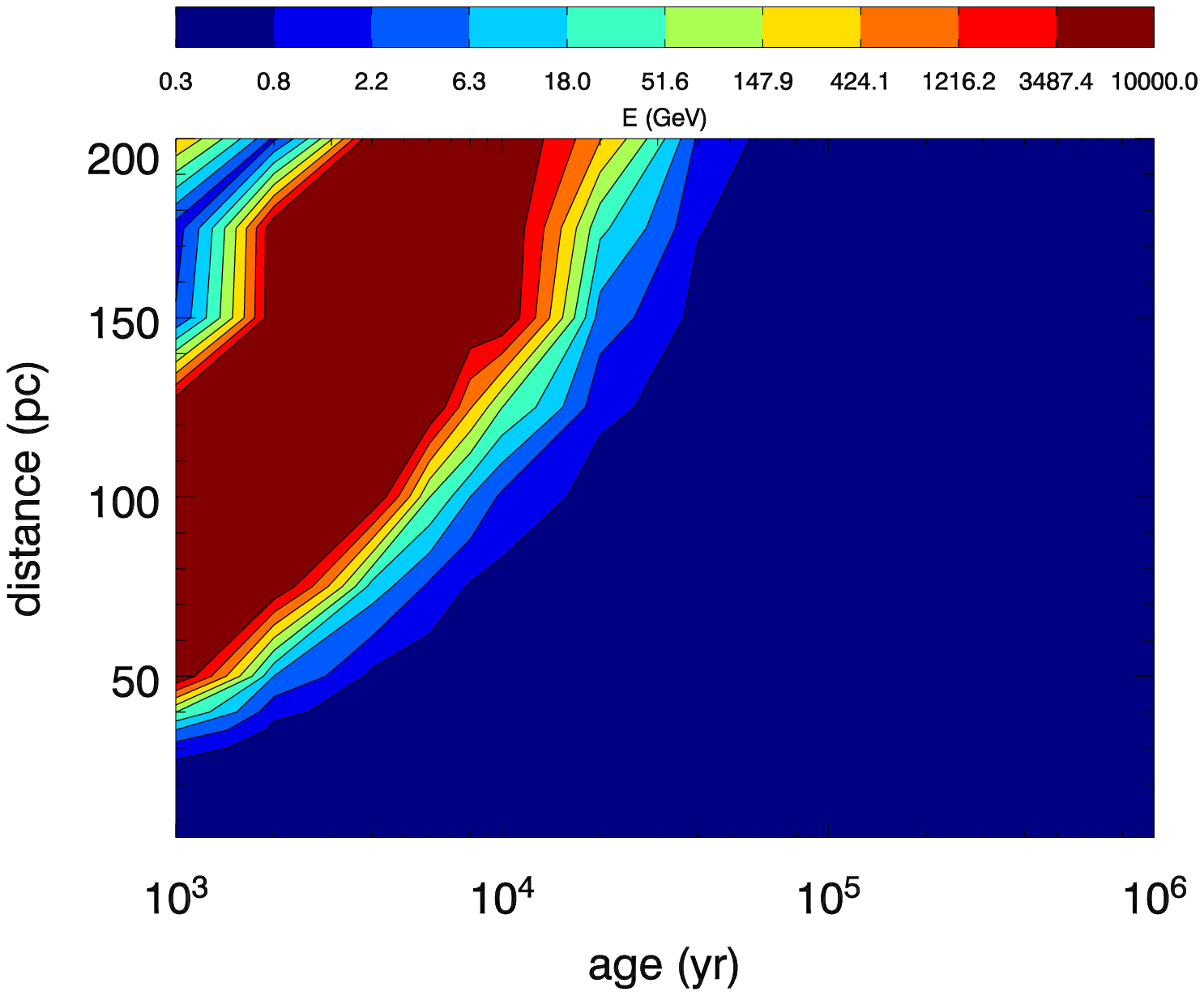}
\includegraphics[width=.71\columnwidth,trim=0 5 0 10]{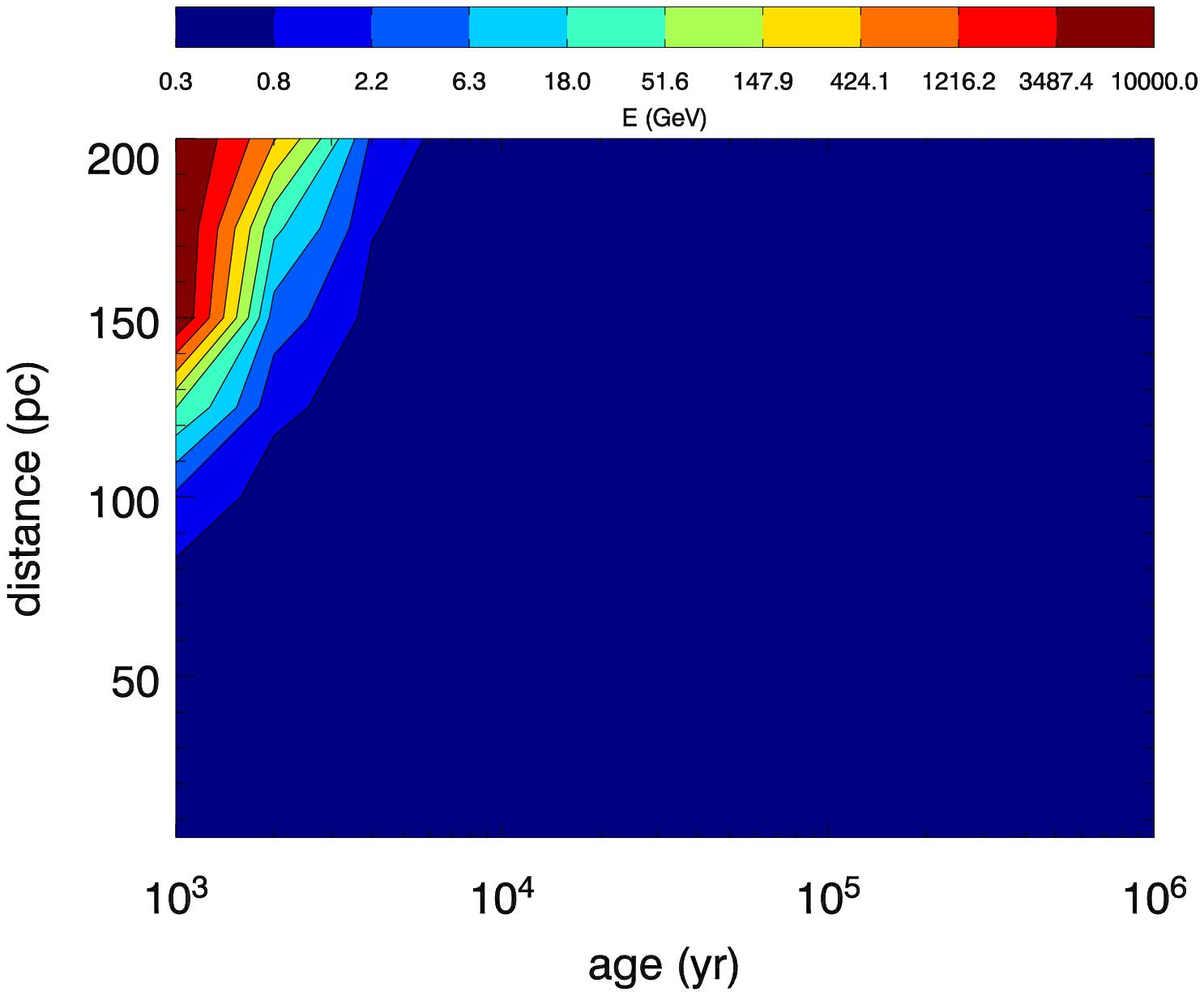}
\caption{
For each combination of age and accelerator-target separation, for which more than two thousand spectra where numerically produced,  the energy of the maximum of such spectra are shown in a contour plot. The color of the different contours corresponds to the range of energy where the maximum is found according to the color bar on the top of the figures. From left to right, plots are created for the case of an impulsive source injecting protons in a medium with $D_{10}$=$10^{26}$cm$^2$/s, $10^{27}$cm$^2$/s 
and $10^{28}$cm$^2$/s. After Rodriguez et al. (2008). }
\label{obs-1}
\end{figure*}

In the framework of diffusion models, Figure  \ref{obs-1} shows, as contour plots, the energy at which the maximum of the SED is to be found for the cases of an impulsive acceleration of cosmic rays, at different distances, ages of the accelerator, and diffusion coefficient. The diffusion radius,  for times less than the proton-proton interaction timescale, $t \ll \tau_{pp}$, is $R_{\rm dif} (E) = 2 \sqrt{D(E) t}$, so that at a fixed age and distance, only particles of higher energy will be able to compensate a smaller $D_{10}$, producing SED maxima at higher $E$-values.   
Small values of $D_{10}$ are expected in dense regions of ISM.
It is interesting to note that for many, albeit not for all, of the SEDs, the maximum in $E^2F$ space is found at energies beyond the energetic range of Fermi, opening thus the gate for studies with the next generation of instruments, particularly observing in the few tens of GeV. 
An  observational discovery of a 1 -- 100 GeV maximum, provides interesting clues about the nature of the astrophysical system that generates the gamma-rays: the range of accelerator-target separations and ages of accelerator that would produce such a 1--100  GeV maximum is rather limited (see in Figure  \ref{obs-1} the narrow contours for maxima at such energies), which would help to produce a direct identification of the source, in case  such system is found in the vicinity of the detection.

\subsection{Molecular clouds}

Another related issue concerns the study of cosmic-ray diffusive penetration into molecular clouds. If the diffusion coefficient inside a cloud is significantly smaller as compared to the average one in the neighborhood, 
low 
energy cosmic-rays are excluded from penetrating deep in the cloud, and part of the gamma-ray emission from the cloud is suppressed, with the consequence of its 
gamma ray spectrum appearing harder than the cosmic-ray spectrum (see e.g., Gabici et al. 2007).
Both of this effects are to be more pronounced in the 
densest central region of the cloud. Thus, an angular resolution of the order of 1 arcmin or less would
allow to resolve the inner part of the clouds and measure the degree of penetration of cosmic-rays into them, 
allowing to put constraints on the change of the diffusion coefficient.

In addition, future instruments will be sensitive enough as to detect
even with no cosmic-ray enhancement, a population of close
(100-300 pc) small clouds, most of them translucent.\footnote{Small molecular clouds are divided
into three classes on the basis of their visual extinction and
astrochemical properties (van Dishoeck \& Black 1988): diffuse,
translucent, and dark. Translucent clouds, the intermediate regime,
have CO abundances in the range $10^{-6}-10^{-4}$ and column
densities greater than $10^{15}$ cm$^{-2}$. Different
from dark clouds, where the chemistry is driven by collisional
processes, translucent clouds are still dominated by photoprocesses.
It is in this translucent regime where most of the carbon becomes
molecular, and so they are detectable in CO, but are relatively
optically thin -- and thus their name -- exhibiting low optical
extinctions. } 
These were discovered in recent 
 sensitive CO surveys at relatively high Galactic latitudes, e.g., covering $|b| < 30$ deg and $\delta >$ -17 deg (l < 230 deg) with a sampling 
interval of 1/4 deg or better (Dame \& Thaddeus 2004). There are about 
 200 relatively small and isolated molecular clouds $|b|>10$ degrees, with 
 masses between 1 and 100 solar. Many of these will be detectable by future-ACTs, what will allow us to feedback upon the mass 
estimation of each of the clouds, and/or  the variation of cosmic-ray spectrum in the Earth vicinity (Torres et al. 2005).

\subsection{Galactic Center}

Having said all above, it is straightforward  that the Galactic Center itself (as discussed in more detail in other contributions to this volume) will be one of the prime science targets 
for the next generation of VHE instruments. Essentially, following the results by Aharonian et al. 2004, 2006b; Albert et al. 2006b),
one can realize of the need of better sensitivity and angular resolution to be able to pursue an 
analysis of cosmic-ray diffusion in the region, with a few pc binning (implying about 1 arcmin of angular separation),
distinguishing between one or multiple originators of the primary cosmic-ray population. 

Indeed, apart from the candidate counterparts found in the error circles of the measured detections of the central Galactic Center source observed both by MAGIC and HESS: the
shell-type SNR Sgr~A~East,
the pulsar wind nebula G\,359.95$-$0.04, 
and the supermassive black hole Sgr~A$^{\star}$ itself; as well as the nearby TeV source coincident with the SNR G\,0.9+0.1, it is interesting to focus on what lies beneath.
The diffuse Galactic center emission turned out to be visible after subtraction of the previous two TeV sources, as reported by Aharonian et al. (2006b). These diffuse emission presented two main features: it appears correlated with molecular clouds (as for instance, traced by the CS (1--0) line), and it by far exceeds (by a factor of 3 to 9) the gamma-ray emission that would be produced if the same target material would be subject to the local neighborhood cosmic-ray environment. 
The observed morphology and spectrum of the gamma-ray emission 
provide evidence that one or more cosmic-ray accelerators have been active in
the Galactic center  in the last 10000 years. These accelerator, then, could provide the cosmic-rays that after hadronic interactions produce the diffuse emission seen. 
An alternative (which future instruments should 
rule out (or not), is to entertain the possibility for a number of Inverse Compton sources to be behind the emission (albeit the reasons for a correlation with the molecular clouds would not be that clear in such a case).

The ratio of gamma-ray emission to molecular density varies with
galactic longitude in a uniform way, except for a pronounced dip 
at $l\approx1.3^{\circ}$ in the TeV emission that is not mimicked by the molecular content (Aharonian et al. 2006b).  The implied non-uniform cosmic ray density
can be explained if cosmic rays injected close to Sgr A have not yet had time
to diffuse out to the $l\approx1.3^{\circ}$ region. Assuming a fixed value for the diffusion coefficient of $D = 10^{30} \ \rm cm^{2} s^{-1}$, what is
typical for the Galactic
Disk at TeV energies, an angular displacement of $1^{\circ}$ at the
distance of the Galactic Center corresponds to a diffusion time of $\sim$10$^{4}$
years, close to the age of the SNR Sgr~A~East. Future instruments with enhanced sensitivity, angular resolution,  and energy coverage would allow to go deeper into this cosmic-ray propagation issue, and actually determine the energy dependent diffusion coefficient observationally (e.g., Hinton et al. 2006). 
A larger sensitivity would also allow similar kind of studies to be done in other Galactic regions, closer to Earth, where the cosmic ray environment is probably less enhanced, and that are also beyond the reach of current IACTs.

\subsection{PWN}

As it is well known, the Crab Nebula is a very effective accelerator (exemplified from observing radiation across more than 15 decades of energy) but not an effective inverse Compton gamma-ray emitter. Indeed, we see gamma-rays from Crab because of its large spin-down reservoir ($\sim 10^{38}$ erg s$^{-1}$), although the gamma-ray luminosity  is much less than its spin-down power, what can be understood because of a large magnetic field, whose strength also depends on the spin-down reservoir. A less powerful pulsar would imply a  weaker magnetic field, what would imply a higher gamma-ray efficiency (i.e., a more efficient sharing between synchrotron and inverse Compton losses). For instance, HESSJ1825-137 (Aharonian el al 2006c) presents a similar luminosity to Crab, but has 2 order of magnitude less spin down power, and its magnetic field has been constrained to be in the range of  a few $\mu$G, instead of hundreds.
The differential gamma-ray spectrum of the whole emission region from this latter object is measured over more than two orders of magnitude, from 270 GeV to 35 TeV, and shows indications for a deviation from a pure power law that future arrays could observe. 
Forecasting future observations, 
spectra have also been determined for spatially separated regions of HESS J1825-137. The photon indices from a power-law fit in the different regions show a softening of the spectrum with increasing distance from the pulsar and therefore an energy dependent morphology. If the emission is inverse Compton generated, the pulsar power is not enough to generate the gamma-ray luminosity, suggesting that the pulsar had a higher injection power in past. Is this common for other PWN?
What can that teach us about the evolution of pulsar winds?
Another notable case is that of 
Vela X (Aharonian el al 2006d), for which the first detection of what appears to be a VHE Compton peak in the SED diagram was found. Although a hadronic interpretation has also been put forward (Horns et al. 2006)  it is yet unclear how large the content of ions in  the pulsar wind could be.  

All in all, the study of pulsar wind nebulae (PWN) with future arrays opens the possibility to focus on several issues related to pulsar winds, and how they reflect on their capability of cosmic-ray accelerators: magnetization, composition, bulk Lorentz factor, 
anisotropy, particle distribution\footnote{See more about this in  the section about binaries.}.

\subsection{Absorption by the local ISRF}

Given the known existence of multi-TeV sources, 
future arrays could also measure VHE absorption in the interstellar radiation field (ISRF). This is in fact impossible for other experiments, like Fermi, due to insufficient energy coverage, and very hard or impossible for current experiments, due to insufficient sensitivity. 
The attenuation due to the CMB is at only 10 Kpc if $E>$500 TeV. But the attenuation due to the ISRF (which has a comparable number density at longitudes 20$\mu$m to 300$\mu$m, can produce absorption at about 50 TeV (Zhang et al. 2006, Moskalenko et al. 2006). Observation on the cutoff energy of different sources will provide independent information to test and constrain the ISRF model. 

\section{Microquasars and gamma-ray binaries}

Very recently, a few massive binaries have been identified as variable VHE  $\gamma$-ray sources. They are
PSR B1259-63 (Aharonian et al. 2005a), LS 5039 (Aharonian et al. 2005b, 2006), LS I +61 303 
(Albert et al. 2006, 2008a,b), and Cyg X-1 (Albert et al. 2007). 
The nature of only two of these binaries is considered known: PSR B 1259-63 is formed with a pulsar whereas Cyg X-1 is formed with a black hole compact object. The nature of the two remaining systems is under discussion. 

The high-energy phenomenology of 
Cyg X-1 is different from that of the others. It has been detected just once in a flare state for which a duty cycle is yet unknown. The three other sources, instead, present a behavior that is fully correlated with the orbital period. The latter varies from about 4 days in the case of LS 5039 to several years in the case of PSR B1259-63. 
Cyg X-1 and the the remaining TeV sources also differ in their SEDs. In Cyg X-1, the transient VHE luminosity was less than 1\% of the X-ray luminosity. For PSR B1259-63, LS 5039 and 
LSI +61 303, the emission above 10 MeV dominates the radiative output.
Are these differences indicative of a universal behavior related with the system composition?

\begin{figure}
  \centering
    \includegraphics[height=.478\textheight]{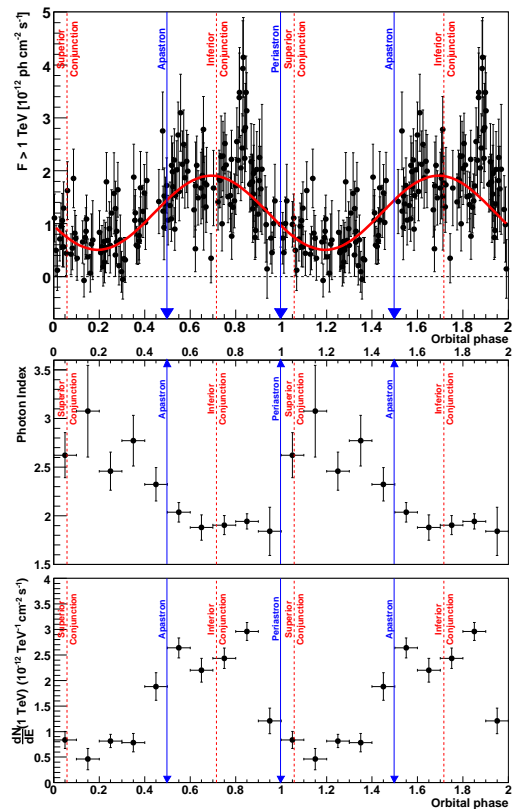}
  \caption{Lightcurve for LS 5039, together with power-law fittings in 0.1-phase intervals. From Aharonian et al. (2006).}
  \label{fig:lightcurves}
\end{figure}

Figure \ref{fig:lightcurves}, from Aharonian et al. (2006), where the reader can also check the spectral variability, presents the current public data on LS 5039. It is one of the richest datasets of high energy astrophysics, and certainly in the top tier of Galactic ones. In the top of the Figure, the $\gamma$-ray flux ($F>1$~TeV) lightcurve (phasogram) of LS~5039 from HESS data (2004 to 2005) is presented on 
on a run-by-run basis. Each run is $\sim$28~minutes.
Two full phase periods are shown for clarity. The blue solid arrows depict periastron and apastron.
The thin red dashed lines represent the superior and inferior conjunctions of the compact object, 
and the thick red line depicts the Lomb-Scargle Sine coefficients for the
period giving the highest Lomb-Scargle power in the periodicity analysis, resulting in the TeV determination of the orbital period. These data has been binned in 0.1-phase intervals, and subsequently fitted with a power law, which normalization and index are shown in the middle and bottom panel of the figure (for energies 0.2 to 5~TeV).  Because of low statistics, and presumably too, of vastly different statistics at the higher end of the spectrum in each of these bins, 
more complicated functions such as a power-law with 
exponential cutoff provide a no better than a pure power-law. This is then a concept for showing a powerful testing machinery: the more sensitive 
the instrument, the smaller the phase bin (the interval) upon which an spectrum can be given. Detailed models should then aim in providing spectral evolution along this and others binary orbits, taking into account the most complete possible physical (e.g., geometrical effects, anisotropic fields, cascading).

The referred question about what kind of systems are these, now remains open. Even, it might be possible to say
that it has been worsened since the Swift-BAT detection of a SGR/AXP-like burst from the direction of LS I+61 303  (see the GCN circular 8209 and their follow ups). The ascription of this short 0.2 s-burst to 
LS I +61 303 itself, what would qualify it most likely as a magnetar (see e.g., ATEL 1715)
has been put to question by several authors (see ATELs 1730, 1731, 1740). Recent reports of observations  
made by RXTE have also called to surprise, finding that the X-ray emission from LS I +61 303 is changing by up to a factor of six over timescales of several hundred seconds as well as doubling in times as fast as 2 seconds (Smith et al. 2008). We recall, however, that RXTE has a large position uncertainty, and since no follow-up observations to any of these flares were made, it is perfectly possible that they were generated in a source nearby (in sky projection) but different from LS I+61 303.

We are not sure whether the identification of sources such as LS 5039 or LS I +61 303 will come before or after
the existence of a new generation of ground based gamma-ray instruments, despite current efforts. The ultimate proof of a pulsar in these systems, if such exist, i.e., the detection of pulsed emission at any wavelength, might remain obscured by strong absorption (see e.g., Zdiarski et al. 2008). If so, in order to confirm or rule out a particular composition for these or other systems, it might be necessary to recur to high energy astrophysics, and compare the most detailed predictions possible with observations having enough statistics in short time-intervals. 

If gamma-ray binaries are pulsars, is the gamma-ray emission coming mostly from processes within the pulsar wind zone or rather from those being produced by particles accelerated in the wind collision shock? Is this answer a function of energy? Gamma-ray astronomy can indeed provide clues not only to answer this question, but also, to acquire information regarding the particle's energy distribution within the pulsar wind zone itself. For instance, data on LS 5039, analyzed in the context of pulsar wind zone models already rule out that mono-energetic electrons are responsible for most of the emission (Sierpowska-Bartosik 2008, Cerruti et al. 2008). What else can future data tell us? How would that impact on current models for dissipation in pulsar winds (see, e.g., Jaroschek et al. 2008 and references therein)? It is not implausible that close systems may trigger different phenomenology within the pulsar wind zone, ultimately affecting particle acceleration there. Models of particle energization and dissipation in pulsar winds are currently made for isolated objects, it is probably true to say that we have no clue whether the inclusion of such objects in close binaries will affect the wind behavior or even the  magnetosphere in appreciable ways. Could we gain knowledge on this using future short-timescales gamma-ray observations?

Among the possibilities for future instruments, it is also worth noticing that of the determination of  the duty cycles of high energy phenomena. For example, 
continued observations of key objects (such as Cyg X-1) with current instrument's sensitivity, using sub-arrays of future ACTs provides a secure science case. A
constant coverage able to measure 0.1 Crab fluxes in less than 1-hour flares, at a Galactic center distance, can be used to measure the duty cycle and provide triggering conditions for the full array of the same TeV instruments, as well as to others at different frequencies (trigger and be trigger by, e.g., LOFAR). We recall that 
energy coverage in 10-100 GeV band, in the case of short flares, is not possible with current instruments (AGILE or Fermi lack sensitivity); nor it is practically possible for higher energies, since that would imply many-months to  year-long campaigns observing the same objects.

Improved angular separation at high energies will also provide new science cases for microquasars, particularly if their jets contain a sizable fraction of relativistic hadrons. While inner engines will still be out of domain for future arrays of high energy telescopes, the study of any possible
microquasar jet/ISM interaction might not be, leading to distinction between the gamma-ray emission produced by the central object (which may be variable) and that produced by the ISM interaction (which may be stable, as in the case of cloud overtaken by a SNR shock), see e.g., Bosch-Ramon et al. (2005). To make this distinction possible,  
at typical kpc distances from Earth, few arcmin from the compact object should be separated.

Finally, also in the case of microquasars, it is known that  
black holes display different X-ray spectral states:
with transitions between a low/hard state, where a compact 
radio jet is found, to a
high/soft state, where the radio emission is quenched by large factors, or there is no detectable radio emission at all (see e.g., Fender 2003).
Are these spectral changes related with
gamma-ray emission?
Is there any
gamma-ray emission during non-thermal radio flares (which can reach up to a
factor of a 1000 increase in radio flux density)?
Indeed,   gamma-ray through inverse Compton
is expected when the radio-to-X-ray
emission during the flares is modeled via synchrotron radiation of relativistic 
electrons suffering radiative, adiabatic and energy-dependent escape losses in
fast-expanding  plasmoids (radio clouds) (see e.g., Aharonian \& Atoyan 1999). Can future gamma-ray observations put constrains upon the  magnetic field  in the plasmoid? Needless to say, the cross-correlations of large-coverage instruments in the radio (e.g., LOFAR) and gamma-rays will be essential in tackling these issues.

\section{Stars and star associations}

O and B stars lose a significant fraction of their mass in stellar
winds with terminal velocities of order $V_\infty=10^3$ km
s$^{-1}$. With mass loss rates as high as
$\dot{M}_\star=(10^{-6}-10^{-4})$ $M_{\odot}$ yr$^{-1}$, the
density at the base of the wind can reach $10^{-12}$ g cm$^{-3}$. 
WR stars represent an evolved stage of hot ($T_\textrm{eff} > 20000$\,K),
massive ($M_\textrm{ZAMS}>25 M_\odot$) stars, and display some of the
strongest sustained winds among galactic objects.  Thus, colliding
winds of massive star binary systems involving WR, O, or B stars are interesting candidates to generate gamma-ray emission.
When two of these massive stars are in binary systems,
the region of the hydrodynamical equilibrium of their winds forms a shock that can accelerate particles: non-thermal spectra in the radio band (e.g., as in the case of Cyg OB2 n.5, or WR 147, e.g., see Williams et al. 1997) have been accordingly detected, as well as hints of correlation of yet unidentified EGRET gamma-ray sources with such binaries have been found (e.g., Romero et al. 1999). 

With different levels of detail, 
 leptonic (inverse Compton of relativistic electrons with the dense photospheric stellar radiation fields in the wind-wind collision zone, e.g., see Reimer et al. 2006 and references therein) and  hadronic (neutral pion decay products, where mesons produced by inelastic interactions of relativistic nucleons with the wind material produce the gamma-rays, e.g., see Benaglia et al. 2001) scenarios have been developed. 
Additionally, inverse Compton pair cascades
initiated by high-energy neutral pion decay photons (from nucleon-nucleon interactions in the stellar winds, e.g., Bednarek 2005) or collective wind scenario in young stellar cluster or OB-associations (e.g., diffusive shock acceleration by encountering multiple shocks, e.g., Klepach et al. 2000;
or inelastic proton interactions with collective winds taking into account the possible convection of cosmic ray primaries, e.g., Torres et al. 2004a, Domingo-Santamaria \& Torres 2006) have been also put forward.  The sensitivity of current  IACTs is not enough to test them thoroughly  and thus these models are being developed 
with basically no observational feedback at the highest energies.

Indeed, sensitive upper bounds upon the gamma-ray emission from isolated binary systems (to avoid the complexity in the determination of origin of the radiation between binaries and collective effects, see the discussion below) were just recently presented by the MAGIC telescope (Albert et al. 2008b). 
Thus the establishment of WR binaries as VHE $\gamma$-ray sources
is yet pending.
The MAGIC
observations of the system WR 147 show already that Fermi should see a flux
cutoff well within its range of detectability in the tens of GeV
regime, if it is able to detect these stars at all. Within the system parameters mentioned
due to the low expected opacity to $\gamma$-ray escape, MAGIC observations would directly limit the acceleration efficiency in this system so as not to produce a significant number of particles with energies typically greater than 100 GeV. 
The next generation of instruments will allow us to go from limits to detection, with the consequent possibility of testing physical assumptions.

Since the HEGRA measurements of the Cygnus OB2 neigborhood and the detection of TeV J2032+4130 (Aharonian et al. 2005), which is still unidentified (Albert et al. 2008), little doubt was left that there exists a direct connection between high-energy photon emission and regions of star formation as a whole.
The observational connection referred to has recently been more firmly established with the detection of the Wd2 cluster (Aharonian et al. 2007). This contains a population of at least 8 stars earlier then O7, 2 WRs, and in particular WR20a, the most massive measured star (81 solar masses) in our Galaxy (a WN6+WN6 binary). This H.E.S.S. source, that shows no variability nor periodicity and is extended, is also still unidentified: it is not yet known whether the emission is produced by a single or by several cluster member (say WR20a and others) or by all, in collective effects within the cluster. Unless there are extreme 
differences in the spatial extent of the particle distributions producing radio, X-ray, and VHE 
gamma-ray emission, scenarios based on the colliding stellar winds in the WR 20a binary system face the problem of accounting for a source extension of 0.2 degrees in the VHE waveband.
But despite this, 
unless an orbital period is detected in future datasets, what in fact will not generically be possibly due to the fact that most expected TeV emitters are long, multi-year period binaries, it would
be very difficult or impossible for the current generation of
instruments to distinguish whether the radiation observed from these
associations (Wd1 is an example, but there is no reason to suppose Wd2, 30 Dor in the LMC, and other associations will not be detected as well) is coming from a few isolated binary systems or is
generated as a collective effect of the whole cluster. Rather, the way to make this distinction lies perhaps in the combined power of an improved angular resolution and sensitivity of future instruments, which will allow to produce detailed maps of gamma-ray emission at different energy cuts in order to compare with theoretical predictions.

Finally, future instruments can aim constructing an overall understanding of the relationship between star formation processes 
and gamma-ray emission. They can experimentally establish whether there is a direct correlation between star formation rate and gamma-ray luminosity when convection and absorption processes at the different environments are taken into account. If such correlation exist, how does it change with energy? Indeed, the discussed stellar associations as Galactic candidates for very high energy emission immediately find extragalactic counterparts in starbursts and luminous infrared galaxies. For the latter, we currently have  only 
upper limits for the nearest starburst (NGC253, Aharonian et al. 2005c) and the nearest ultra-luminous infrared galaxy (Arp 220, the place in the universe with the highest supernova explosion rate known, Albert et al. 2007c). These galaxies are expected to appear in gamma-rays provided enough sensitivity is achieved.
Detailed models have been constructed for NGC 253 which could be tested already with the current instruments (e.g., Domingo-Santamaria \& Torres 2005); others,  like for instance the case of Arp 220 (Torres 2004) require yet further sensitivity.

\section{Concluding remark}

Galactic physics will blossom with a new-generation of instruments like the planned CTA, and after the pathmaking findings of the Fermi survey. It will
enter into an era of precision high-energy observations, it will not be just a discovery trip, as it has partly  been  up to now, but a real immersion into understanding the physical processes generating the emission detected 
and into detailed model testing.


\begin{theacknowledgments}
This work was supported by grants AYA 2006-00530 and CSIC-PIE 200750I029.
\end{theacknowledgments}


\end{document}